\newcommand*{\ATLASLATEXPATH}{}
\documentclass[texmf,cernpreprint,USenglish,texlive=2011]{\ATLASLATEXPATH atlasdoc}
\usepackage{\ATLASLATEXPATH atlaspackage}
\usepackage{\ATLASLATEXPATH atlasbiblatex}
\usepackage{\ATLASLATEXPATH atlascontribute}
\usepackage{\ATLASLATEXPATH atlasphysics}
\graphicspath{{}{}}
\def\Zjets{ \ensuremath{Z/\gamma^{*}{\rm+jets}} }

\AtlasTitle{Search for the standard model Higgs boson produced in association with a vector boson and decaying into a tau pair in $pp$ collisions at $\sqrt{s} = 8$ TeV with the ATLAS detector}

\author{The ATLAS Collaboration}

\PreprintIdNumber{CERN-PH-2015-226}

\AtlasJournalRef{Phys. Rev. D 93, 092005 (2016)}
\AtlasDOI{10.1103/PhysRevD.93.092005}

\AtlasAbstract{%
 A search for the standard model Higgs boson produced in association with a vector boson with the decay $H\rightarrow \tau\tau$ is presented.  The data correspond to 20.3 fb$^{-1}$ of integrated luminosity from proton-proton collisions at $\sqrt{s} = 8$ TeV recorded by the ATLAS experiment at the LHC during 2012.   The data agree with the background expectation, and 95\% confidence-level upper limits are placed on the cross section of this process.  The observed (expected) limit, expressed in terms of the signal strength $\mu = \sigma/\sigma_{\mathrm{SM}}$ for $m_\mathrm{H} = 125 \GeV$, is $\mu <$ 5.6 (3.7).  The measured value of the signal strength is $\mu = 2.3\pm1.6$.
}

\hypersetup{pdftitle={ATLAS draft},pdfauthor={The ATLAS Collaboration}}

\begin{document}

\maketitle

\section{Introduction}
\label{sec:intro}

The investigation of the origin of electroweak symmetry breaking and the experimental confirmation of the Brout--Englert--Higgs mechanism \cite{Englert:1964et,Higgs:1964ia,Higgs:1964pj,Guralnik:1964eu,Higgs:1966ev,Kibble:1967sv} is one of the primary goals of the physics program at the Large Hadron Collider (LHC)~\cite{1748-0221-3-08-S08001}. With the discovery of a Higgs boson with a mass of $125 \GeV$ by the ATLAS \cite{paper2012ichep} and CMS \cite{CMSpaper} collaborations, an important milestone has been reached. To date, measurements of the couplings of the discovered particle \cite{Aad:2013wqa,Aad:2015vsa,CMSFinalCouplings, Chatrchyan:2014vua} as well as tests of the spin--parity quantum numbers ~\cite{Aad:2015mxa,Chatrchyan:2012jja,Khachatryan:2014kca} are consistent with the predictions for the standard model (SM) Higgs boson.

In this paper, a search for the associated production of the Higgs boson with a vector boson, where the Higgs boson decays to a pair of tau leptons, is presented. This production mechanism is referred to in the following as $VH$,  where $V$ is either a $W$ or $Z$ boson.  The analysis is part of a comprehensive program by the ATLAS Collaboration at the LHC to measure the Higgs boson production mechanisms, its couplings, and other characteristics.  Similar studies have been performed with the $VH$ production mechanism and subsequent decays of the Higgs boson to $WW^*$~\cite{Aad:2015ona,CMSWW} and $b\bar{b}$~\cite{Aad:2014xzb,CMS:VHbb} by the ATLAS and CMS Collaborations and to tau lepton pairs~\cite{CMStautau} by the CMS Collaboration.  The associated production is particularly useful in the decays of the Higgs boson to tau lepton pairs when both tau leptons decay hadronically, where the trigger can be a challenge.  For $VH$ production and leptonic decays of the $W$ or $Z$ boson, the $W$ and $Z$ boson decay products satisfy the trigger requirements with high efficiency.

$VH \rightarrow W$/$Z \tau\tau$ production results in several different final-state signatures, which are exploited by an event categorization designed  to achieve both a good signal-to-background ratio and good resolution for the reconstructed $H \to \tau^+\tau^-$ invariant mass. 
Signatures consistent with $ZH$ and $WH$ production are exploited, where only the $W \to \ell \nu$ and the $Z \to \ell\ell$ decays are considered, with $\ell = e, \mu$.  The $H  \to \tau^+\tau^-$ decay signal is reconstructed in the following two possible final states:  both tau leptons decay to hadrons and a neutrino ($\tau_{\text{had}}\tau_{\text{had}}$), or one tau lepton decays leptonically ($\tau \to \ell \nu\bar{\nu}$) and one to hadron(s) and a neutrino ($\tau_{\text{lep}}\tau_{\text{had}}$).

\section{ATLAS detector and object reconstruction}
\label{sec:detector}

The ATLAS detector \cite{DetectorPaper:2008} is a multipurpose detector with a cylindrical geometry.\footnote{The ATLAS experiment uses a right-handed coordinate system with its origin at the nominal interaction point (IP) in the center of the detector and the $z$-axis along the beam direction. The $x$-axis points from the IP to the center of the LHC ring, and the $y$-axis points upward. Cylindrical coordinates $(r,\phi)$ are used in the transverse $(x,y)$ plane, $\phi$ being the azimuthal angle around the beam direction. The pseudorapidity is defined in terms of the polar angle $\theta$ as $\eta=-\ln\tan(\theta/2)$. The angular distance $\Delta{}R$ in the $\eta$--$\phi$ space is defined as $\Delta{}R=\sqrt{({\Delta\eta})^2 + ({\Delta\phi})^2}$.}   It consists of three subsystems: an inner detector (ID) surrounded by a thin superconducting solenoid, a calorimeter system, and a muon spectrometer in a toroidal magnetic field. 

The ID tracking system reconstructs the trajectory of charged particles  in the pseudorapidity range |$\eta$| < 2.5.  It enables the accurate determination of charged-particle momentum and the position of $b$-hadron decay vertices. The inner detector is built from three concentric detector systems surrounded by a solenoid providing a uniform axial 2 T field. The three detector systems are  the pixel detector, the silicon microstrip detector, and the transition radiation tracker. 

The ID tracking system is surrounded by high-granularity lead/liquid-argon (LAr) sampling electromagnetic calorimeters covering the pseudorapidity range |$\eta$|  < 3.2.  A steel/scintillator tile calorimeter provides hadronic energy measurements in the  pseudorapidity region |$\eta$|  < 1.7. In the  regions 1.5 < |$\eta$| < 4.9, the hadronic energy measurements  are provided by two endcap LAr calorimeters using copper or tungsten as absorbers. 
 
The muon spectrometer surrounds the calorimeters. It extends tracking beyond the calorimeter, which enables the identification of muons and a precision measurement of their properties.  It consists of three large superconducting eight-coil toroids, a system of tracking chambers, and detectors for triggering.  Muon tracking is performed with monitored drift tubes covering |$\eta$|  < 2.7 and cathode strip chambers covering |$\eta$|  > 2.0, while trigger information is collected in the resistive plate chambers in the barrel (|$\eta$| < 1.05)   and thin-gap chambers in the endcap regions (1.05 <|$\eta$| < 2.4).
 
A three-level trigger system \cite{TriggerPapaer:2012} is used to select events. A hardware-based level-1 trigger uses a subset of detector information to reduce the event rate to a value of 75 kHz or less. The rate of accepted events is then reduced to about 400 Hz by two software-based trigger levels, level-2 and the event filter. 

A primary vertex is identified for each event.  The reconstructed primary vertex position \cite{ATLAS:2010lca} is required to be consistent with the interaction region and to have at least five associated tracks with transverse momentum $\pT  > 400 \MeV$; 
when more than one such vertex is found, the vertex with the largest summed $\pT^2$ of the associated tracks is chosen.

The tau leptons that decay to hadron(s) and a neutrino, or $\tau_{\text{had}}$, are reconstructed using clusters of energy deposited in the electromagnetic and hadronic calorimeters that are matched to tracks in the inner detector. The identification algorithm separates $\tau_{\text{had}}$ candidates from jets using $\tau_{\text{had}}$ decay characteristics, namely the number of tracks, the collimation of energy deposits in the calorimeter,  and the mass of the $\tau_{\text{had}}$ candidate.  The analysis presented here utilizes $\tau_{\text{had}}$ candidates seeded by an anti-$k_t$ jet algorithm with radius parameter R=0.4 \cite{Cacciari:2005hq, Cacciari:2008gp}, with jet $\pt >$ 20 GeV and |$\eta$| < 2.5.  The $\tau_{\text{had}}$ candidates must have only one or three associated tracks in a cone of size $\Delta R =$ 0.2.  All $\tau_{\text{had}}$  candidates are required to have charge $\pm$1, calculated by summing the charges of the associated tracks. The $\tau_{\text{had}}$ decay products are identified by a  boosted decision tree (BDT) \cite{ATLAS_Tauperf}, which returns a number between zero and one depending on how jet-like or tau-like the reconstructed object is.  The BDT selects taus  with a 55--60\% efficiency (medium $\tau_{\text{had}}$ identification) depending on the $\tau_{\text{had}}$ number of tracks, $\eta$, and $\pt$.  
Dedicated algorithms reject candidates originating from electrons and muons.

Electron candidates are reconstructed from clusters of energy deposited in the electromagnetic calorimeter that are matched to tracks in the inner detector.  They are required to be within the pseudorapidity range |$\eta$| < 2.47 and must have shower shape and track measurements that fulfill the set of medium quality criteria \cite{ATLAS_electrons}, which provides electron identification efficiencies of 80-90\% depending on the transverse energy $\ET$, and $\eta$ of the electron candidate.  Electrons are considered isolated based on tracking and calorimeter information.  
The calorimeter isolation requires
the sum of the transverse energy in the calorimeter in a
cone of size $\Delta{}R$  = 0.4 around the electron cluster, divided
by the $\ET$ of the electron cluster, to be less than 8\% of the electron cluster $\ET$.
The track-based isolation requires the sum of the
transverse momenta of tracks within a cone of $\Delta{}R$ = 0.2 around the electron, divided by the $\ET$ of the electron cluster, to be less than 8\% of the electron cluster $\ET$.

Muon candidates are reconstructed from tracks in the inner detector matched to tracks in the muon spectrometer. 
A requirement on the distance between the primary vertex and the point where the muon candidate track crosses the beamline reduces the background from cosmic rays and beam-induced backgrounds.  Muon candidates are required to be within the pseudorapidity range |$\eta$| < 2.5 and must satisfy a set of quality criteria \cite{ATLAS_muons}, which provides muon identification efficiencies above 95\%.  Muons are considered isolated based on tracking and calorimeter information with similar requirements as are used for electrons, with the muon track $\pT$ in place of the electron cluster $\ET$.
 
Jets are reconstructed from clusters in the calorimeter using the anti-$k_t$ $R = 0.4$ jet algorithm.  Corrections for the detector response are applied   \cite{Cacciari:2007fd, ATLASJets2}. To reduce the contamination of jets by additional interactions in the same or neighboring bunch crossings (pileup), tracks originating from the primary vertex must contribute at least 50\% of the total scalar sum of track $\pT$ within the jets.  This requirement is only applied to jets with $\pT$ <  $50 \GeV$ and |$\eta$| < 2.4.  
 
A $b$-tagging algorithm that relies on tracking information and $b$-hadron characteristics, such as the presence of a decay that can be separated from the primary vertex, is used to identify $b$-jets \cite{ATLASBJets}. The operating point for $b$-tagging chosen for this analysis has a 70\% efficiency for $b$-jets in simulated $\ttbar$ events with a corresponding misidentification probability for light-quark jets of 1\%.  

Missing transverse momentum, with magnitude $\MET$, is reconstructed using the
energy deposits in calorimeter cells calibrated according to the reconstructed physics objects
($e$, $\mu$, $\tau_{\text{had}}$, jets) with which they are associated. Energy deposits not
associated with a physics object tend to have low $\pT$ and are scaled by a dedicated algorithm tuned to improve the resolution
in high-pileup conditions \cite{ATLASMET}.

\section{Data and simulation samples}
\label{sec:data}

The analysis uses those data collected when the detector systems were certified as functioning properly. The resulting data sample corresponds to an integrated luminosity of  $20.3~\ifb$ of pp collisions at $\sqrt{s}=8$ TeV. Samples of signal and background events are simulated using a number of Monte Carlo (MC) generators, listed in Table~\ref{tab:MCGenerator}. The cross-section values to which the simulation is normalized and the perturbative order in quantum chromodynamics (QCD) for each calculation are also provided.  For the signal samples, the central value of the factorization scale equals the sum of the Higgs boson mass and the vector boson mass.

The generated events are combined with minimum-bias events simulated using the \textsc{AU2}~\cite{atlasmctunes} parameter tuning of \textsc{Pythia8}~\cite{Sjostrand:2007gs} to take into account multiple interactions.  All simulated events undergo  full simulation of the ATLAS detector response \cite{Aad:2010ah} using the \textsc{Geant4} \cite{Agostinelli:2002hh}  simulation program before being processed through the same reconstruction algorithms as the data. The signal samples use the CTEQ6L1~\cite{CTEQ6L1} PDF set.

\begin{table}[ht]
	\caption{
Monte Carlo generators used to model the signal and the background processes at $\sqrt{s}=8\TeV$. The cross sections times branching fractions ($\sigma \times
	{\cal B}$) used for the normalization of some processes are included in the last column together with
	the perturbative order of the QCD calculation.  For the signal process only the
	$H\to\tau\tau$ SM branching fraction is included. For the $W$ and $Z/\gamma^{*}$
	background processes the branching ratios for leptonic decays ($l = e, \mu, \tau$)  are included.  For all other background processes,
	inclusive cross sections are quoted (marked with a $\dagger$). }
	  \begin{center}
	  \small
	    \begin{tabular}{lllrl}
	    \hline
  {Signal   (Higgs boson mass $m_H=125$~\GeV) }   & {MC generator}  & \multicolumn{3}{l|}{$\sigma \times {\cal B}$ [pb] at  $\sqrt{s}=8~\TeV$} \\
   \hline \hline
 $WH$, $H\to\tau\tau$    &  \textsc{Pythia8}                                                      & 0.0445 & NNLO & \cite{Brein:2003wg,YellowReportIII}\\
 $ZH$, $H\to\tau\tau$    &  \textsc{Pythia8}                                                      & 0.0262 & NNLO & \cite{Brein:2003wg,YellowReportIII}\\
	      \hline \hline
 {Background} & &  \multicolumn{3}{l}{}\\
  \hline \hline
$W (\to l \nu$), ($l= e, \mu, \tau)$                  & \textsc{Alpgen}~\cite{ALPGEN}+\textsc{Pythia8}      & 36800  & NNLO & \cite{PhysRevLett.103.082001,PhysRevLett.98.222002}\\
$Z/\gamma^{*}(\to ll)$,& \textsc{Alpgen+Pythia8}   & 3910 & NNLO & \cite{PhysRevLett.103.082001,PhysRevLett.98.222002}\\
 ~~60~\GeV$<m_{ll}<2$~\TeV   & & & &\\
$Z/\gamma^{*}(\to ll)$,   & \textsc{Alpgen+Herwig}~\cite{herwig}     & {13000}  & {NNLO} & {\cite{PhysRevLett.103.082001,PhysRevLett.98.222002}}\\
	      ~~10~\GeV$<m_{ll}<60$~\GeV  & & & &\\
	      	   	   $t\bar{t}$                         & \textsc{MC@NLO~\cite{MC@NLO} + Jimmy~\cite{Jimmy}}                                          & 238$^{\dagger}$  & NLO & \cite{MC@NLO} \\
	      $q\bar{q} \to WW$                  & \textsc{Alpgen+Herwig}                                           & 54$^{\dagger}$ & NLO & \cite{MCFMVV}\\
	      $gg \to WW $                       & \textsc{gg2WW}~\cite{Binoth:2006mf}+\textsc{Herwig}                       & 1.4$^{\dagger}$ & NLO & \cite{Binoth:2006mf}\\
	      $WZ,ZZ$                            & \textsc{Herwig}                                                  & 30$^{\dagger}$ & NLO & \cite{MCFMVV}\\ 
	      \hline
	    \end{tabular}
	\label{tab:MCGenerator}
	  \end{center}
	\end{table}

\section{Event categorization and selection}
\label{sec:eventselection}

A characteristic of $VH$ production is the presence of a $W$ or $Z$ boson in each signal event.   The analysis categories are optimized to exploit the leptonic decays of the vector bosons that provide a candidate for the electron or muon triggers and to reduce the backgrounds from multijet processes.  The presence of additional leptonic and/or hadronic tau decays from the Higgs boson allows for the event selection to include a requirement on three or four objects, depending on the channel, to define the final state. 

The single-lepton and dilepton triggers  used to select the events in this analysis are listed in Table \ref{Table:TriggerThresholds}. The \pT\ requirements on the particle candidates in the analysis are 2~\GeV\ higher than the trigger thresholds, to ensure that the trigger is maximally efficient.

\begin{table}[htp]
\caption{ Summary of the triggers used to select events for the various channels. The transverse momentum thresholds applied at trigger level are listed.}
\begin{center}
\begin{tabular}{ll}
\hline \hline
Trigger  & Trigger threshold(s) [\GeV]  \\
\hline
Single electron                                & $\pt^{e}>24 $                                       \\

Single muon                                 & $\pt^{\mu}>24 $                            \\ 

Combined electron and muon                              & $\pt^{e}>12$     ~~~~    $\pt^{\mu}>8$       \\

Symmetric dielectron                                     & $\pt^{e_1}>12$        ~~~  $\pt^{e_2}>12$   \\

Asymmetric dielectron                & $\pt^{e_1}>24$    ~~~      $\pt^{e_2}>7$           \\

Symmetric dimuon                                                      & $\pt^{\mu_1}>13$   ~~~    $\pt^{\mu_2}>13$       \\

Asymmetric dimuon                            & $\pt^{\mu_1}>18$     ~~~   $\pt^{\mu_2}>8$     \\
\hline\hline
\end{tabular}
\label{Table:TriggerThresholds}
\end{center}
\end{table}

The four analysis event categories are determined by the type of associated vector boson and the topology of the $H\rightarrow\tau\tau$ decay.  These are summarized in Table \ref{cuts} and described below.   

\begin{table}[ht]
	\caption{Summary of the selection criteria for each of the four analysis channels.}	
	\centering
	\begin{tabular}{lll}
		\hline
		Channel & Selections \\
		\hline
		 $W\rightarrow\mu\nu/e\nu, ~~H\rightarrow\tau_{\text{lep}}\tau_{\text{had}}$ 		& Exactly one isolated electron and one isolated muon\\
													 	& Exactly one $\tau_{\text{had}}$ passing medium BDT ID \\
													 	& $\pT(\tau_{\text{had}}) > 25 \GeV$ \\
													 	& Same-charge $e$ and $\mu$, oppositely charged  $\tau_{\text{had}}$ \\
													 	& Events containing $b$-tagged jets with $\pT > 30 \GeV$ are vetoed \\
													 	& $|\pT(\tau_{\text{had}})| + |\pT(\mu)| + |\pT(e)| > 80 \GeV$ \\
                                                        & $\Delta R(\tau_{\text{had}}, \tau_{\text{lep}}) < 3.2$ \\
		\hline
		$W\rightarrow\mu\nu/e\nu, ~~H\rightarrow\tau_{\text{had}}\tau_{\text{had}}$ 	& Exactly one isolated electron or one isolated muon \\
																& Exactly two $\tau_{\text{had}}$ passing medium BDT ID of opposite charge \\
																& $\pT(\tau_{\text{had}}) > 20 \GeV$ \\
                                              
													 	        & $|\pT(\tau_{\text{had}}^1)| + |\pT(\tau_{\text{had}}^2)| > 100 \GeV$ \\
                                                                & $m_{\mathrm{T}}(\ell, \MET) > 20\GeV$ \\
                                                                & $0.8 < \Delta R(\tau_{\text{had}}^1, \tau_{\text{had}}^2) < 2.8$ \\
													 			& Events containing $b$-tagged jets with $\pT > 30 \GeV$ are vetoed \\
		\hline
	$Z\rightarrow\mu\mu/ee, ~~H\rightarrow\tau_{\text{lep}}\tau_{\text{had}}$	& Exactly three electrons or muons,\\
															& One opposite-charge and same-flavor lepton pair  \\
															& with invariant mass $80 < m_{\ell\ell} < 100 \GeV$ \\
															& Exactly one $\tau_{\text{had}}$ passing medium BDT ID, with opposite charge \\
															& to the lepton assigned to the Higgs boson  \\
															& $\pT(\tau_{\text{had}}) > 20 \GeV$ \\
													 	    & $|\pT(\tau_{\text{had}})| + |\pT(\tau_{\text{lep}})| > 60 \GeV$ \\
		\hline
 $Z\rightarrow\mu\mu/ee, ~~H\rightarrow\tau_{\text{had}}\tau_{\text{had}}$	& Exactly two electrons or two muons of opposite charge\\
														& Exactly two $\tau_{\text{had}}$ passing medium BDT ID of opposite charge\\
														& $\pT(\tau_{\text{had}}) > 20 \GeV$ \\
														& $ 60 < m_{\ell\ell} < 120 \GeV $ \\
														& $ |\pT(\tau_{\text{had}}^1)| + |\pT(\tau_{\text{had}}^2)| > 88 \GeV$ \\
		\hline
	\end{tabular}
	\label{cuts}
\end{table}

{\bf The $W\rightarrow\mu\nu/e\nu, ~~H\rightarrow\tau_{\text{lep}}\tau_{\text{had}}$ channel}: These events are required to have one isolated electron, one isolated muon, and one $\tau_{\text{had}}$ candidate.  The electron and muon candidates  are required to have an electric charge of the same sign to reduce the backgrounds from $\ensuremath{Z/\gamma^{*}\rightarrow{\tau\tau}{\rm+jets}} $ events, $WW$ events, and $t\bar{t}$ events where both $W$ bosons decay leptonically.  The electron or muon  candidate with the higher $\pt$ is assumed to arise from the $W$ boson decay, which is correct 75\% of the time in the MC simulation.  The $\tau_{\text{had}}$ candidate is required to have $\pt > 25 \GeV$ and to have opposite electric charge to the leptons.  Events containing $b$-tagged jets with $\pt>30\GeV$ are vetoed  to further reduce the background from $\ttbar$ events.   The scalar sum of the $\pt$ of the electron, muon, and $\tau_{\text{had}}$ candidates must be greater than $80 \GeV$  to reduce the backgrounds from multijet and $\Zjets$ events.   To reduce backgrounds from quark- or gluon-initiated jets misidentified as $\tau_{\text{had}}$ when these jets are produced back-to-back, the angle between the $\tau_{\text{had}}$ and $\tau_{\text{lep}}$ candidates associated with the Higgs boson is required to satisfy $\Delta R(\tau_{\text{had}}, \tau_{\text{lep}})  < 3.2$.

{\bf The  $W\rightarrow\mu\nu/e\nu, ~H\rightarrow\tau_{\text{had}}\tau_{\text{had}}$ channel}: These events are required to have one isolated electron or muon candidate and two $\tau_{\text{had}}$ candidates.  The two $\tau_{\text{had}}$ candidates are required to have  $\pt > 20 \GeV$  and to have opposite charge.  The lepton is assumed to come from the $W$ boson. Events containing $b$-tagged jets with $\pt>30\GeV$ are vetoed to reduce the background from $\ttbar$ events. The scalar sum of the $\pt$ of the lepton and two $\tau_{\text{had}}$ candidates must be greater than $100 \GeV$  in order to reduce the background from multijet events. The transverse mass\footnote{The transverse mass, $m_{\mathrm{T}}  = \sqrt{2\pt^{\ell}\MET \cdot (1 - \cos\Delta\phi)}$ where $\Delta\phi$ is the azimuthal separation between the directions of the lepton and the missing transverse momentum.  }  of the lepton and $\MET$ must be greater than $20 \GeV$. To reduce the background from events with jets misidentified as $\tau_{\text{had}}$ candidates, $0.8 < \Delta R(\tau_{\text{had}}^1, \tau_{\text{had}}^2) < 2.8$ is required, which results in a reduction of the background from misidentified jets by almost a factor of two while losing less than a third of the signal events.

{\bf The $Z\rightarrow\mu\mu/ee, ~H\rightarrow\tau_{\text{lep}}\tau_{\text{had}}$ channel}: Events containing one $\tau_{\text{had}}$ candidate and three light lepton candidates  are in this category. The two light lepton candidates with invariant mass closest to $91 \GeV$, opposite electric charge, and the same flavor are assumed to be the $Z$ boson decay products.  The invariant mass of the leptons assumed to come from the $Z$ must be between $80$ and $100 \GeV$. The remaining light lepton and the $\tau_{\text{had}}$ candidate are assumed to originate from the Higgs boson decay. They are thus required to have opposite charge and the scalar sum of their $\pt$ values must be greater than $60 \GeV$. 

{\bf The  $Z\rightarrow\mu\mu/ee, ~H\rightarrow\tau_{\text{had}}\tau_{\text{had}}$ channel}:  Signal candidates are selected by requiring exactly two electron (muon) candidates and two  $\tau_{\text{had}}$  candidates. The two light leptons are assigned to the $Z$ boson decay, are required to have the same flavor, and are required to have opposite electric charge.  The invariant mass of the two lepton candidates assigned to the $Z$ boson must be  between $60$ and $120 \GeV$. The two $\tau_{\text{had}}$  candidates are assumed to originate from the Higgs boson decay and are required to have opposite electric charge.  A minimum requirement of $88 \GeV$  is placed on the scalar sum of the transverse momenta of the $\tau_{\text{had}}$ pair to reduce the
$\Zjets$ background. 

After all the analysis selection criteria are applied, the number of events migrating  from other Higgs boson channels, in particular from $VH$ production where the Higgs boson decays into $WW$, is found to be negligible.  This analysis selection has an acceptance of 1.9\% for the combined $WH$ channels, where the denominator requires a light lepton from the $W$ boson decay ($W\rightarrow\mu\nu/e\nu/\tau_{e/\mu}\nu$) and for the Higgs boson to decay through the considered tau decay chains ($H\rightarrow\tau_{\text{lep}}\tau_{\text{had}}$ or $H\rightarrow\tau_{\text{had}}\tau_{\text{had}}$), and the numerator includes all analysis cuts.  The acceptance for the combined $ZH$ channels is 5.3\%, where the denominator requires a light lepton pair from the $Z$ boson decay ($Z\rightarrow\mu\mu/ee/\tau\tau_{\mu\mu/ee}$) and for the Higgs boson to decay through the considered tau decay chains ($H\rightarrow\tau_{\text{lep}}\tau_{\text{had}}$ or $H\rightarrow\tau_{\text{had}}\tau_{\text{had}}$), and the numerator includes all analysis cuts.

\section{Background estimation}
\label{sec:backgrounds}

The number of expected background events and the associated kinematic distributions are derived using data-driven methods as well as simulation. 
There are two classes of backgrounds for this analysis: processes in which all three or four final-state lepton and $\tau_{\text{had}}$ candidates are actually produced, and those in which some lepton or $\tau_{\text{had}}$ candidates  are actually misidentified jets.
Jets are most likely to be misidentified as $\tau_{\text{had}}$ objects, although the rate at which jets mimic electrons is, in some instances, not negligible.

Backgrounds containing real electrons, muons, and $\tau_{\text{had}}$ leptons primarily arise from  diboson, $Z\to\tau\tau$, and $\ttbar$ events. These backgrounds are determined from Monte Carlo simulation. The background arising from jets misidentified as electron or $\tau_{\text{had}}$ candidates is estimated using a data-driven method, the so-called fake-factor method. The $\tau_{\text{had}}$ fake factor is defined as the ratio of the number of $\tau_{\text{had}}$  candidates identified with medium $\tau_{\text{had}}$ criteria to the number satisfying the loosened but not the medium identification criteria. The electron fake factor is defined as the number of electrons satisfying the identification criteria divided by the number of those that do not.  The fake-factor measurements are described below.  For the $W\rightarrow\mu\nu/e\nu, ~H\rightarrow\tau_{\text{lep}}\tau_{\text{had}}$ channel both the $\tau_{\text{had}}$ and electron fake factors are used, while for the other three channels the $\tau_{\text{had}}$ fake-factor method alone performs well enough for modeling the background from misidentified jets.  The background from misidentified jets is the dominant background, or comparable to the background from diboson production, in all channels of the analysis. 

Since the fake rates are sensitive to the underlying physics of the event, the fake factors are measured in a region with similar kinematics and composition of misidentified objects to the signal region.  Applying the analysis selection to MC simulation reveals that $\Zjets$ events are the primary source of the background from misidentified jets in the analysis. The rate of jets mimicking the $\tau_{\text{had}}$  selection is therefore measured using a tag-and-probe method from jets in well-reconstructed $Z/\gamma^{*}\rightarrow\mu\mu$+jets events. The tag here is the di-muon system and the probe is the additional jet(s) that may be suitably tau-like (pass medium $\tau_{\text{had}}$  identification) or suitably jet-like (pass a loosened $\tau_{\text{had}}$  identification but fail the medium one). 
 The fake factor is measured as a function of the jet $\pT$, $\eta$, and number of associated tracks.
The fake rate for electrons is calculated separately, using well-reconstructed  $Z \to \mu\mu$ events containing additional jets or photons, using the same procedure as described above. 

To estimate the background from misidentified jets for the $WH$ and $ZH$ signal regions, these factors are then applied to the event combinations that have all selections the same as the signal selection with the exception that at least one $\tau_{\text{had}}$  candidate has passed the loosened but failed the medium $\tau_{\text{had}}$  identification.  For the $W\rightarrow\mu\nu/e\nu, ~H\rightarrow\tau_{\text{lep}}\tau_{\text{had}}$ channel, a contribution from jets misidentified as the electron candidate is also taken into account using objects that have failed electron identification.  Since many background events contain multiple jets that could potentially pass the $\tau_{\text{had}}$ or electron identification, more than one possible combination of passing and failing objects is allowed to contribute per event. In these cases, the multiple copies of the events contribute with the various weights calculated for each combination of objects considered.

The fake-factor method is validated independently in each of the four analysis channels.  In each case a comparison between the data and the background prediction is made with a loosened signal selection, which provides a test of the method with a large number of events in a dataset that is dominated by the background from misidentified jets.  In addition, a series of orthogonal regions are formed to validate the method for each of the analysis channels.  The definition of the loosened signal selection and validation regions are given for each channel in Table~\ref{Table:FFValidation}. 

\begin{table}[ht]
	\begin{center}
    \caption{The loosened signal selection and the list of validation regions used to validate the fake-factor method are given for each of the four analysis channels.  MMC and $M_{\mathrm{2T}}$ are mass reconstruction techniques defined in Section 6.}
	\small
	\begin{tabular}{lll}
	    \hline
	    Channel                                                                     & Loosened signal selection & Validation regions \\
	    \hline
	    $W\rightarrow\mu\nu/e\nu, ~H\rightarrow\tau_{\text{lep}}\tau_{\text{had}}$ & one isolated electron &  $Z \rightarrow\tau\tau$: $Z$ mass selection (60--120$\GeV$)\\
	    															&  one isolated muon & $\ttbar$: require $b$-tagged jet \\
													 			& $\pT(\tau_{\text{had}}) > 25 \GeV$ &  \\
																\hline
        $W\rightarrow\mu\nu/e\nu, ~H\rightarrow\tau_{\text{had}}\tau_{\text{had}}$ & one isolated electron or muon & $Z \rightarrow\tau\tau$: $Z$ mass selection (>60$\GeV$)  \\
        															& two (opposite charge) $\tau_{\text{had}}$ candidates &  $\ttbar$: require $b$-tagged jet \\
        															&  & $W$ + jets: $m_{\mathrm{T} }(\ell, \MET) > 60\GeV$ \\
        															&  &  same-sign $\tau_{\text{had}}$ candidates \\
        															&  & mass sideband: $M_{\mathrm{2T}}$< 60$\GeV$ \\
															&  & or $M_{\mathrm{2T}}$>120$\GeV$ \\	
        														\hline																			
        $Z\rightarrow\mu\mu/ee, ~H\rightarrow\tau_{\text{lep}}\tau_{\text{had}}$   & three isolated electrons or muons  & same-sign $\tau_{\text{lep}}$, $\tau_{\text{had}}$ candidates \\       					
        															& opposite-charge, same-flavor lepton pair & mass sideband: $M_{\mathrm{MMC}}$<80$\GeV$  \\
															&  $\tau_{\text{had}}$ with opposite charge to the  & or $M_{\mathrm{MMC}}$>120$\GeV$  \\
															& unpaired lepton &  \\
        \hline
        $Z\rightarrow\mu\mu/ee, ~H\rightarrow\tau_{\text{had}}\tau_{\text{had}}$  & two opposite-charge, same-flavor leptons  & same-sign $\tau_{\text{had}}$ candidates \\        															
        															& two opposite-charge $\tau_{\text{had}}$ candidates & mass sideband: $M_{\mathrm{MMC}}$<80$\GeV$\\
															& &  or $M_{\mathrm{MMC}}$>$120\GeV$ \\

	    \hline
	\end{tabular}
	\label{Table:FFValidation}
	\end{center}
\end{table}

Example distributions of the $\pT$ of $\tau_{\text{had}}$ candidates for the loosened signal selection and validation regions are shown in Figure \ref{Figure:FFValidation} for the $W\rightarrow\mu\nu/e\nu, ~H\rightarrow\tau_{\text{lep}}\tau_{\text{had}}$ channel.  MC simulation studies show that this $Z\rightarrow\tau\tau$ validation region is dominated by $Z\rightarrow\tau\tau$ events where an additional jet in the event is misidentified as a $\tau_{\text{had}}$ candidate.  Likewise, MC simulation studies show that this $\ttbar$ validation region is dominated by $\ttbar$ events where at least one $W$ boson decays leptonically and where a jet is misidentified as a $\tau_{\text{had}}$ candidate. The number of expected signal events and estimated total number of background events for each channel in the signal region are given in Table \ref{results}. 

\begin{figure}[ht]
        \centering
           \subfloat{
   \includegraphics[width=0.49\textwidth]{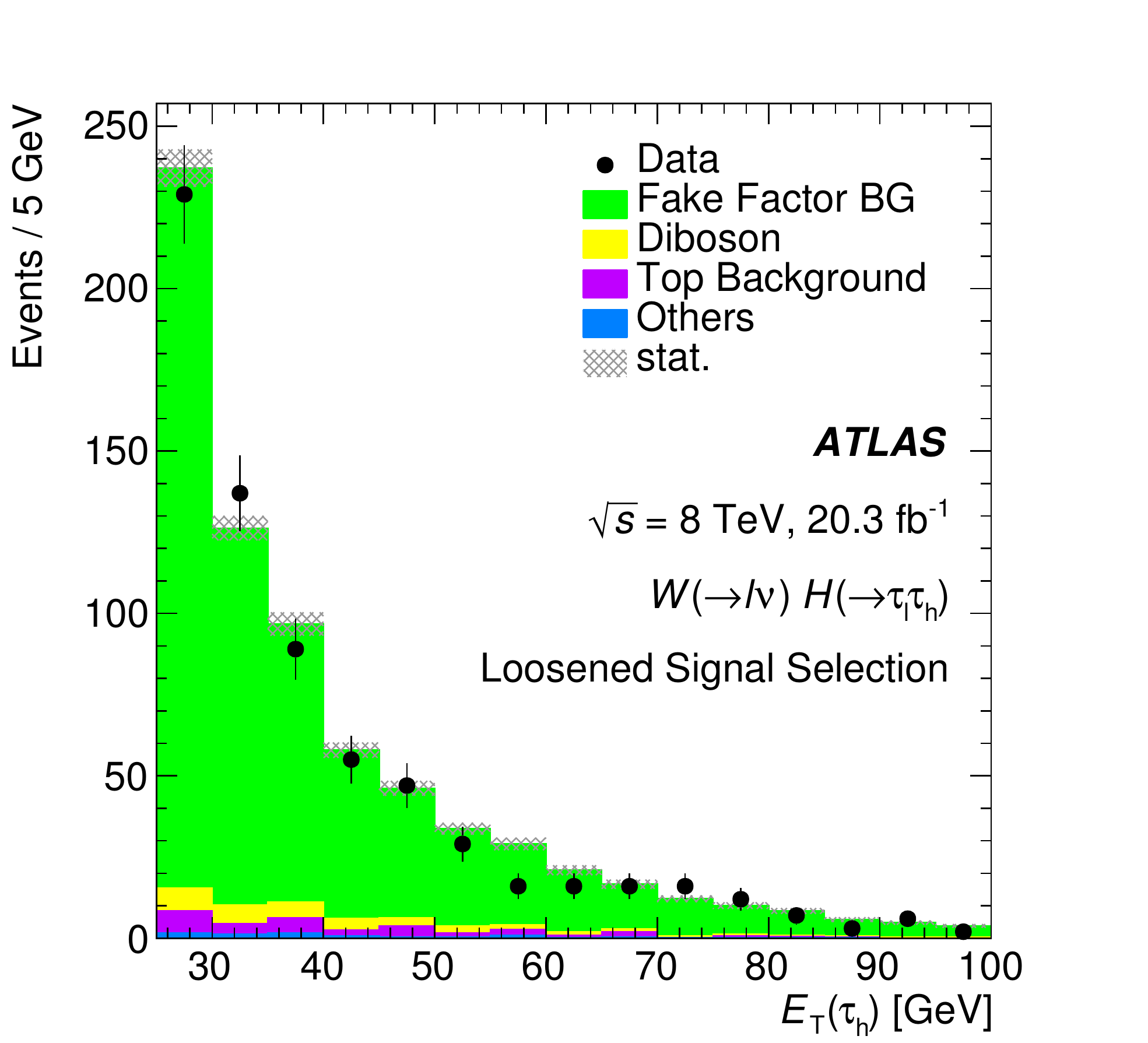}
          } 
             \subfloat{
     \includegraphics[width= 0.49\textwidth]{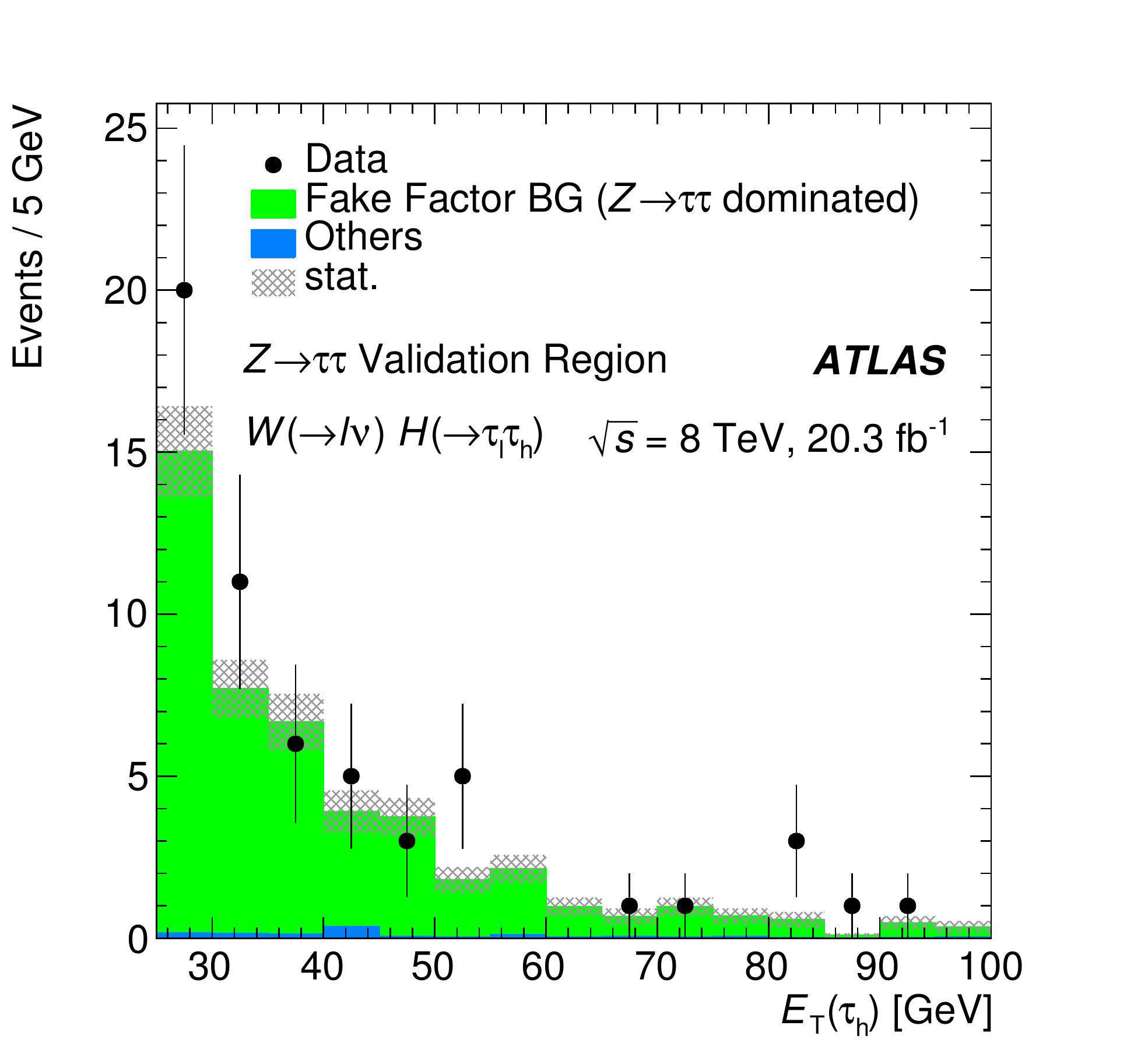}
          }\\
             \subfloat{
     \includegraphics[width= 0.49\textwidth]{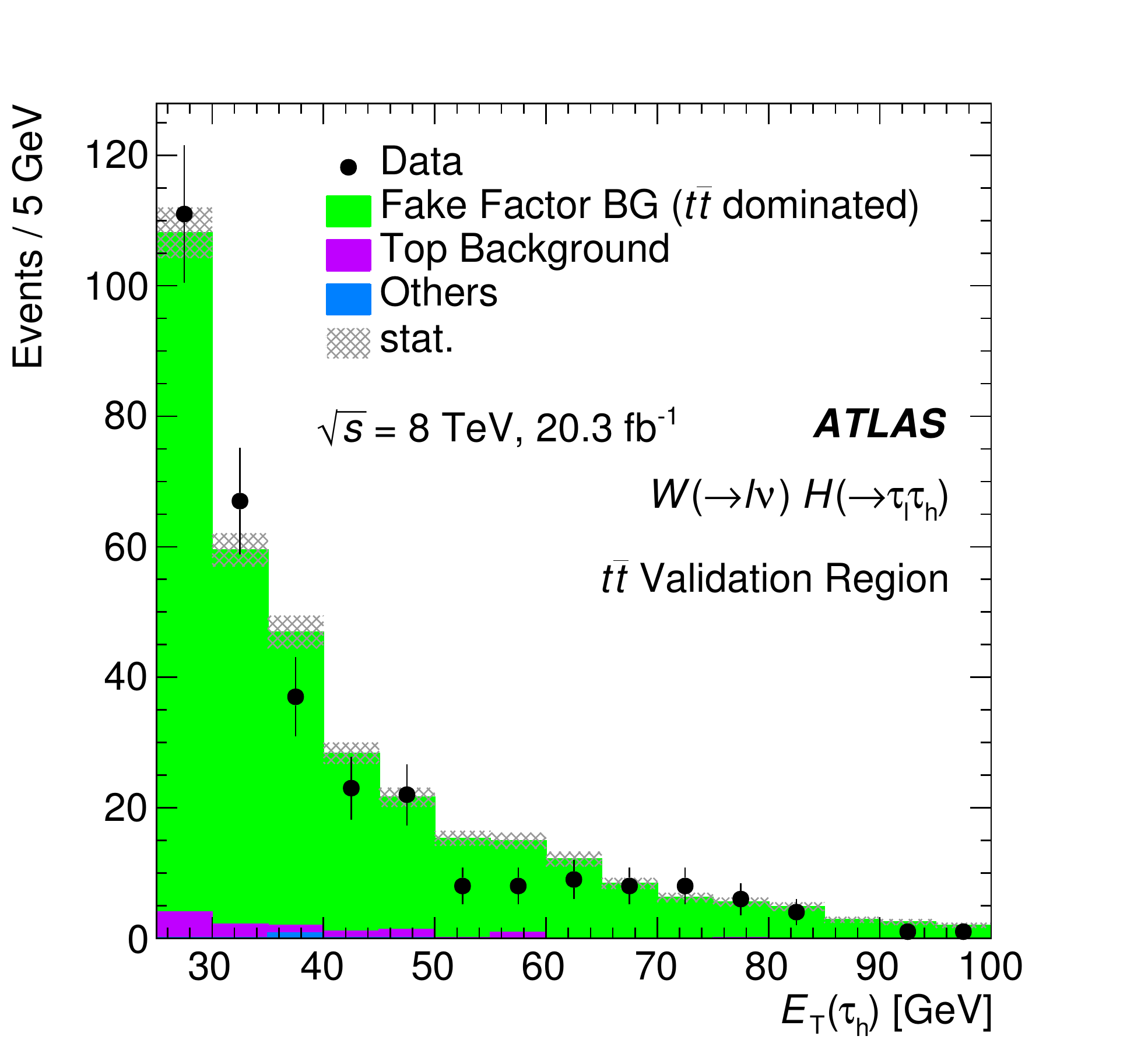}
          }
          
           \caption{Distributions of the transverse energy of the $\tau_{\text{had}}$
candidate, as validation for the fake factor method for the $W\rightarrow\mu\nu/e\nu, ~H\rightarrow\tau_{\text{lep}}\tau_{\text{had}}$ channel.  The category labeled "Fake Factor BG" consists of events where at least one $\tau_{\text{had}}$ or electron candidate does not result from a simulated $\tau_{\text{had}}$ or electron. }
         \label{Figure:FFValidation}
\end{figure}

\begin{table}[ht]
	\begin{center}
    \caption{The yields for the observed and expected background and signal for a 125 GeV Higgs boson in the signal region for each individual channel.  The "other" column consists primarily of background from $\ttbar$ events. The uncertainties quoted are statistical only.}
	\small
	\begin{tabular}{lccclll}
	    \hline
	    Channel                                                                     & Obs. & Signal        & $\Sigma$ Background & Fake Factor & Diboson & Other \\
	    \hline
	    $W\rightarrow\mu\nu/e\nu, ~~H\rightarrow\tau_{\text{lep}}\tau_{\text{had}}$ & 35  & $1.95\pm0.05$ & $32.4\pm1.9$ & $13.1\pm1.3$ & $13.54\pm0.35$ & $5.7\pm1.4$ \\
        $W\rightarrow\mu\nu/e\nu, ~~H\rightarrow\tau_{\text{had}}\tau_{\text{had}}$ & 33  & $1.84\pm0.04$ & $35.5\pm2.7$ & $28.1\pm2.4$ & $7.4\pm1.2$ & - \\
        $Z\rightarrow\mu\mu/ee, ~~H\rightarrow\tau_{\text{lep}}\tau_{\text{had}}$   & 24  & $1.14\pm0.03$ & $24.6\pm1.5$ & $17.1\pm1.5$ & $7.28\pm0.16$ & $0.20\pm0.01$ \\
        $Z\rightarrow\mu\mu/ee, ~~H\rightarrow\tau_{\text{had}}\tau_{\text{had}}$   & 7   & $0.64\pm0.02$ & $6.8\pm1.2$  & $4.7\pm1.2$ & $2.09\pm0.09$ & $0.012\pm0.003$ \\
	    \hline
	\end{tabular}
	\label{results}
	\end{center}
\end{table}

\section{Mass reconstruction}
\label{sec:massreconstruction}

The result is extracted using a fit to the reconstructed invariant mass or transverse mass spectrum of the $\tau_{\text{lep}}$--$\tau_{\text{had}}$ or $\tau_{\text{had}}$--$\tau_{\text{had}}$ pair. The mass is reconstructed using one of two methods, depending on the signal category.  The Higgs boson mass in  $ZH$ events is calculated using  the missing mass calculator (MMC) method described in Ref.  \cite{ATLAS_MMC}. This method takes the $x$- and $y$-components of the event missing transverse momentum as an input as well as the visible  mass of the $\tau_{\text{lep}}$--$\tau_{\text{had}}$ or $\tau_{\text{had}}$--$\tau_{\text{had}}$ pair.  Because the neutrinos from the tau decays have unknown $x$-, $y$- and $z$-components and there are multiple neutrinos (two for the $\tau_{\text{had}}$--$\tau_{\text{had}}$ case and three for the $\tau_{\text{lep}}$--$\tau_{\text{had}}$ case), the system is underconstrained.  A scan is therefore performed over possible momenta for the neutrinos, and a most-likely di-$\tau$ mass is found.  

 In the $WH$ category, the presence of an additional neutrino from the $W$ decay makes the MMC mass reconstruction not optimal. In this case the $M_{\mathrm{2T}}$ variable defined in Ref.\cite{Barr:2011xt} is used, which calculates an event-by-event lower bound (within the detector resolution) of the transverse mass of the $\tau_{\text{had}}$-$\tau_{\text{had}}$ or $\tau_{\text{lep}}$-$\tau_{\text{had}}$ pair by performing a minimization over the allowed phase-space of possible
momenta of assumed neutrinos in the event.  In the general case described in Ref. \cite{Barr:2011xt} the only constraint on the phase-space is that the sum of the transverse momenta
of all neutrinos equals the observed $\MET$.  For this analysis, the additional constraint that the invariant mass of the lepton and neutrino assigned to 
 the $W$ boson be equal to, or as close as possible to, the mass of the $W$ boson is imposed.
 The mass distributions after all the selection criteria are applied are shown in Figure \ref{Figure:M2TANDMMC}. 
\begin{figure}[ht]
        \centering
     \subfloat{
   \includegraphics[width=0.49\textwidth]{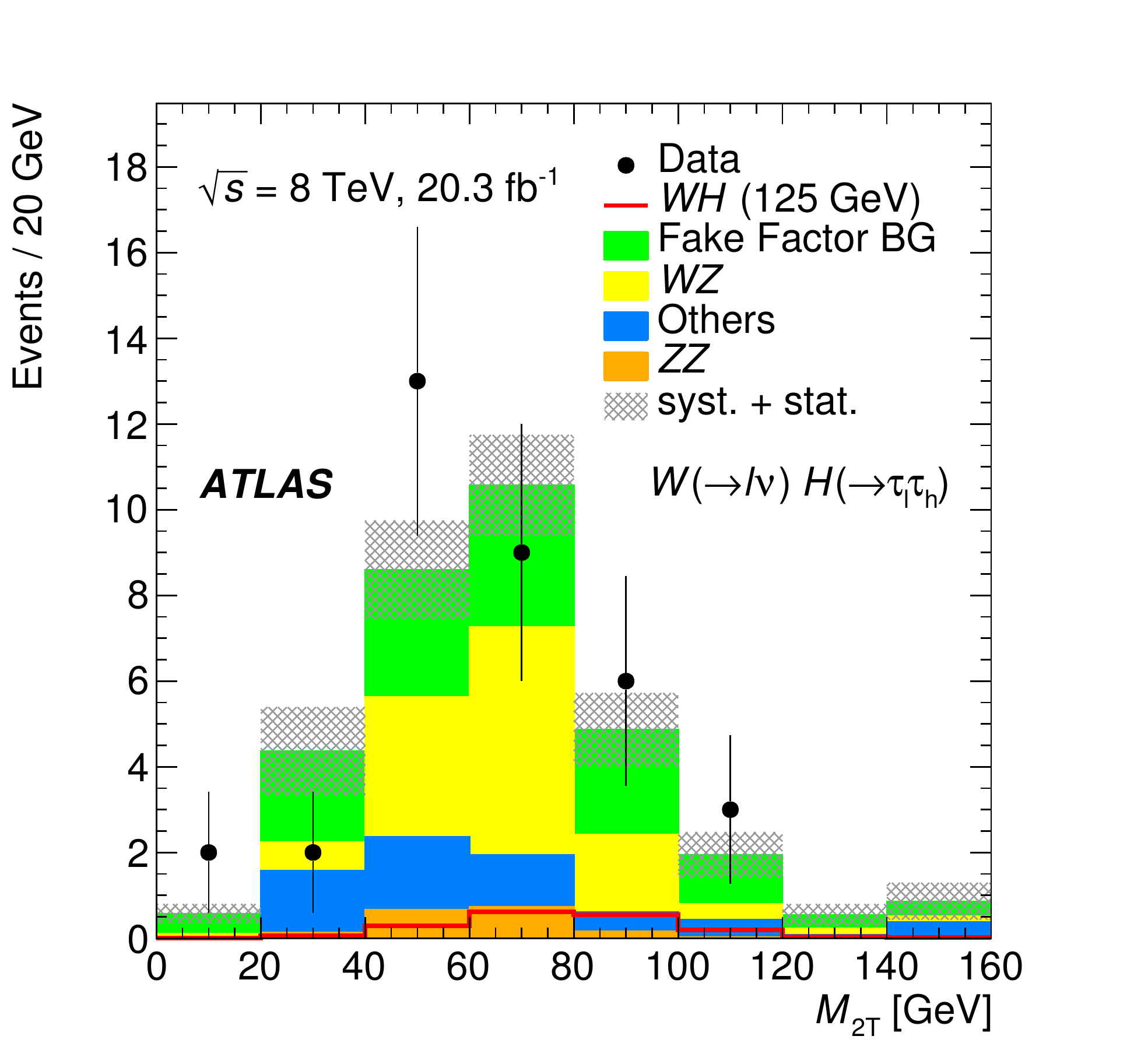}
          } 
            \subfloat{ 
     \includegraphics[width= 0.49\textwidth]{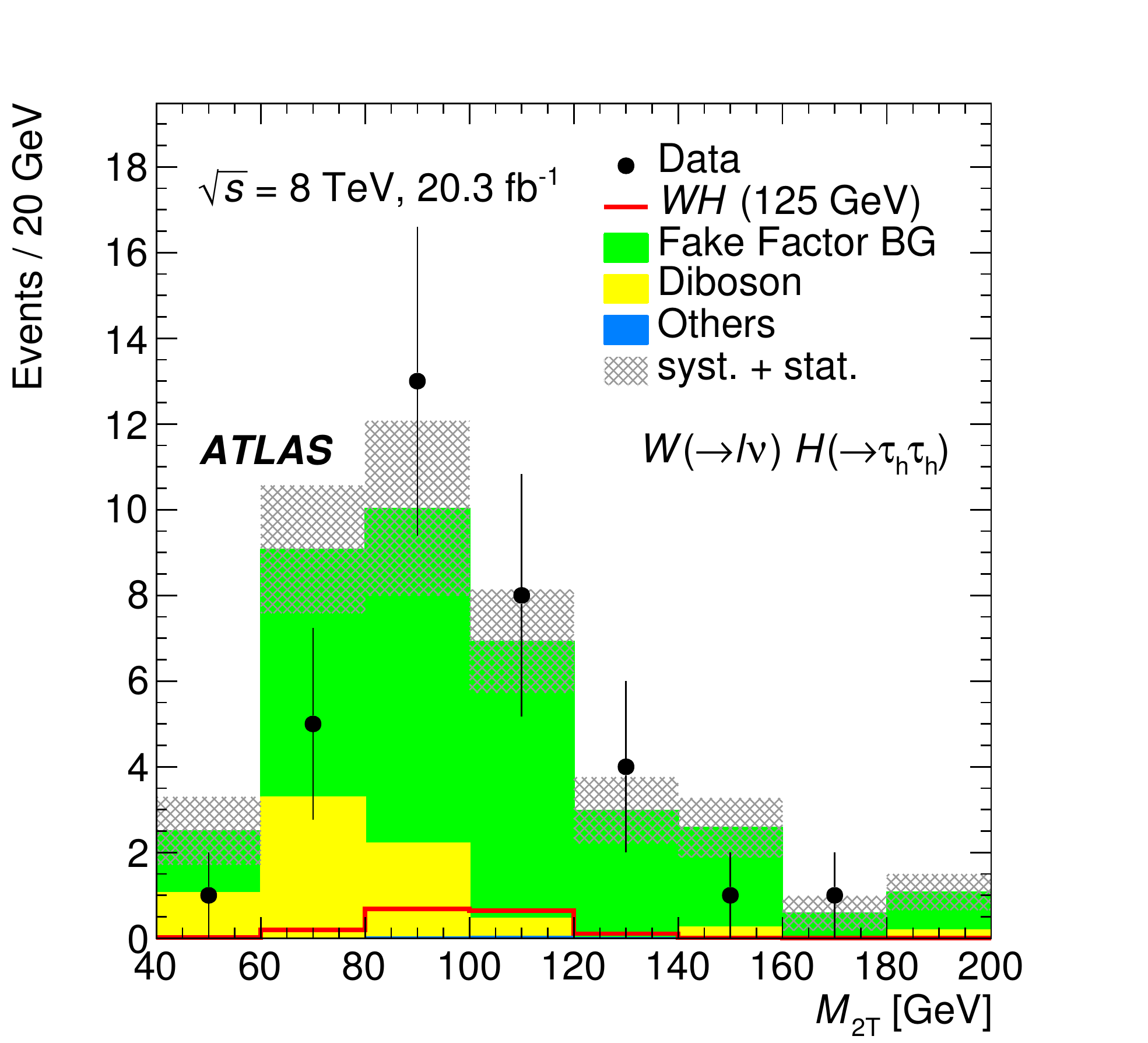}
          }\\
      \subfloat{
     \includegraphics[width= 0.49\textwidth]{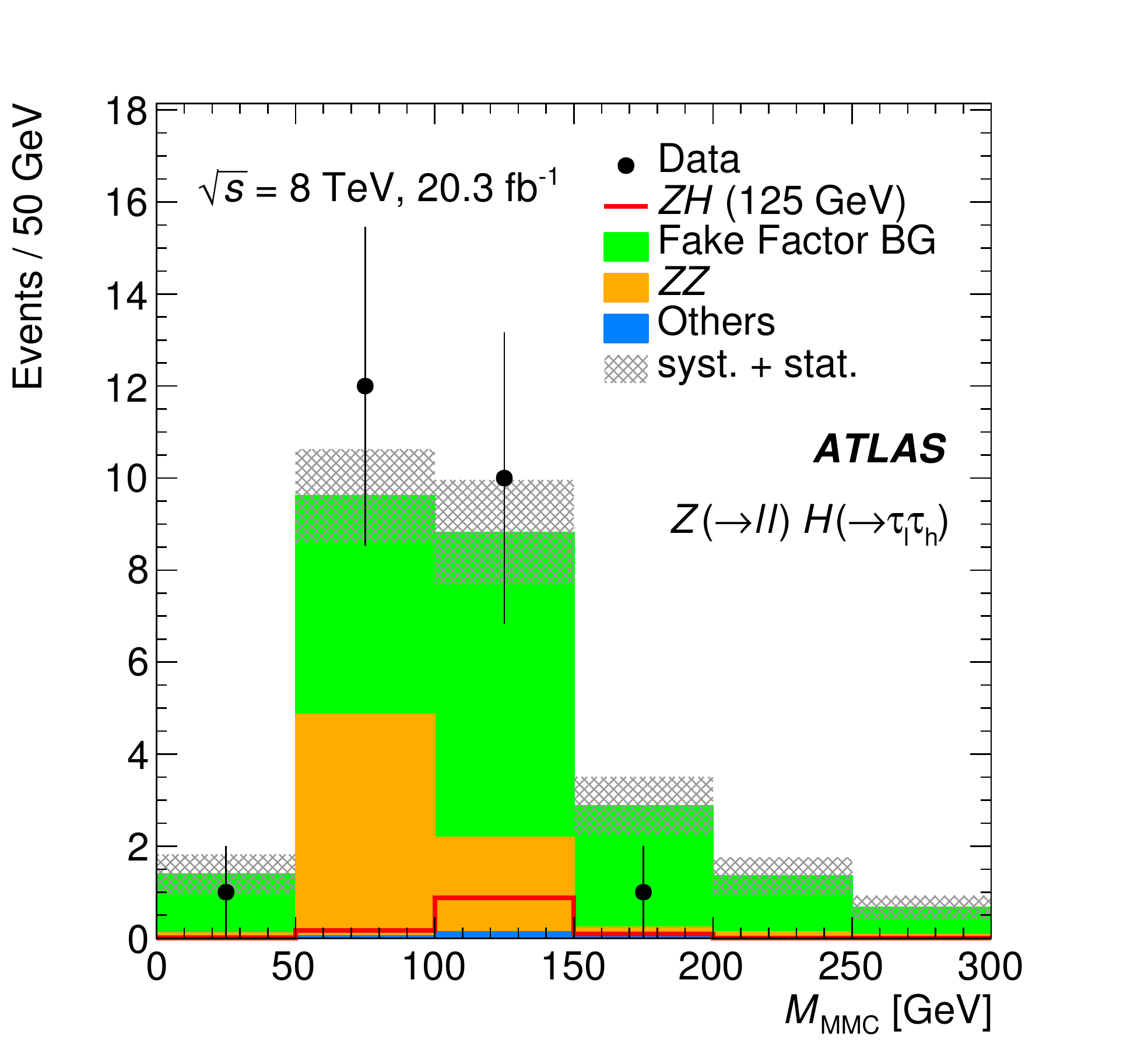}
          }
     \subfloat{
     \includegraphics[width= 0.49\textwidth]{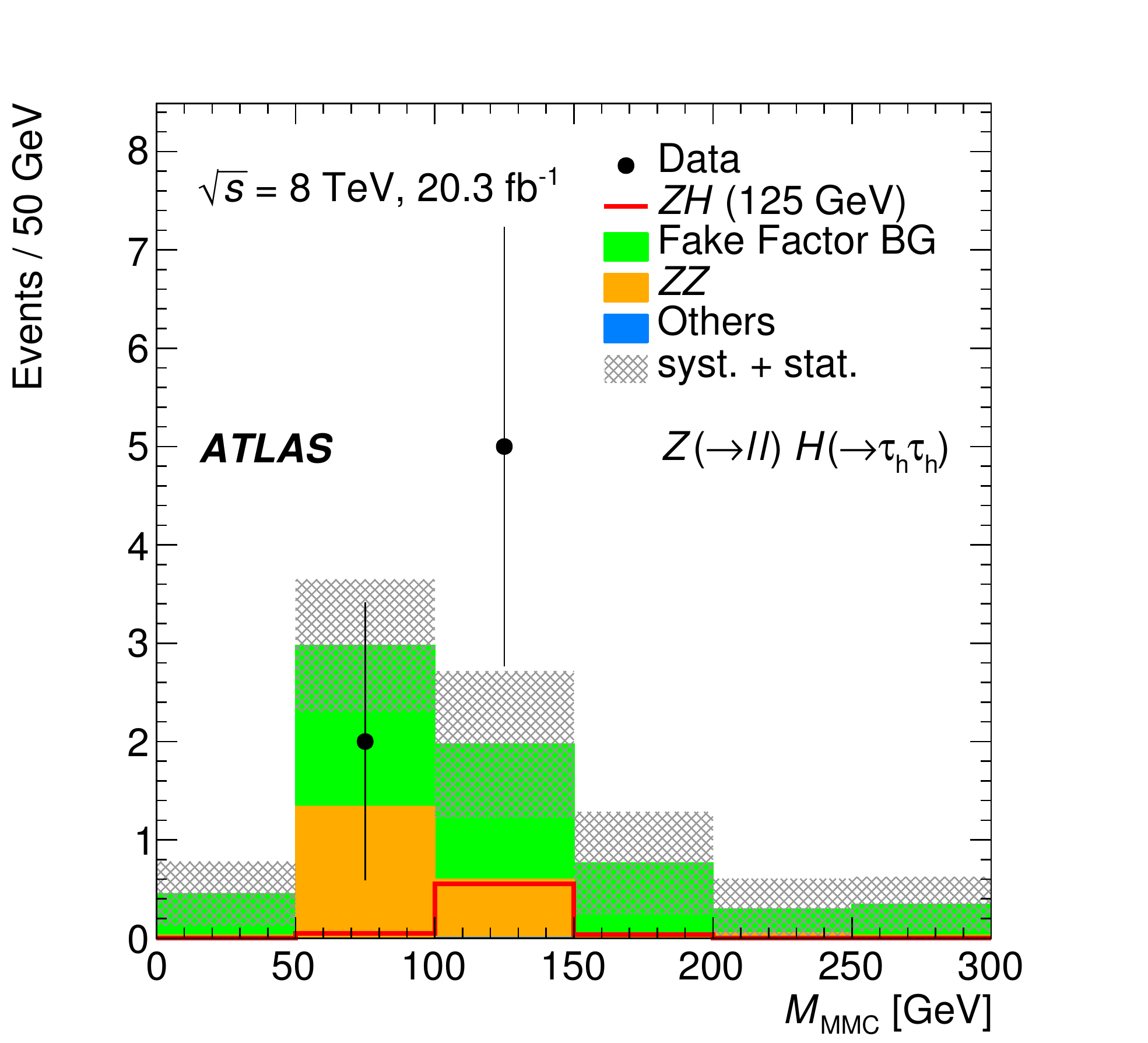}
          }
          
           \caption{Mass distributions used to determine the strength of signal in each channel.  Upper left: $M_{\mathrm{2T}}$ distribution for the $WH\rightarrow\tau_{\text{lep}}\tau_{\text{had}}$ channel.  Upper right: $M_{\mathrm{2T}}$ distribution for the $WH\rightarrow\tau_{\text{had}}\tau_{\text{had}}$ channel.  Lower left: $M_{\mathrm{MMC}}$ distribution for the $ZH\rightarrow\tau_{\text{lep}}\tau_{\text{had}}$ channel.  Lower right: $M_{\mathrm{MMC}}$ distribution for the $ZH\rightarrow\tau_{\text{had}}\tau_{\text{had}}$ channel.}
         \label{Figure:M2TANDMMC}
\end{figure}

\section{Systematic uncertainties}
\label{sec:systematicuncertainties}

The numbers of expected signal and background events, and the distributions of the discriminating variables $M_{\mathrm{MMC}}$ and $M_{\mathrm{2T}}$, are affected by systematic uncertainties. These uncertainties are discussed below and are grouped into three categories: experimental uncertainties, background modeling uncertainties, and theoretical
uncertainties. For all uncertainties, the effects on both the total signal and background yields and on the shape of the mass distributions, $M_{\mathrm{MMC}}$ or  $M_{\mathrm{2T}}$ respectively, are evaluated.  Table \ref{systematics-prefit} shows the systematic uncertainties, their impact on the number of expected events
for the signal and the relevant background, and their impact on the post-fit signal strength, $\mu$, where $\mu = \sigma/\sigma_{\mathrm {SM}}$ and the value ${\cal B}$($H\rightarrow\tautau$) corresponds to the standard model prediction for $m_H$ = $125\GeV$.

Experimental systematic uncertainties arise  from uncertainties on trigger efficiencies, particle reconstruction and identification,  uncertainties on the energy scale and resolution of jets, leptons, and $\tau_{\text{had}}$ candidates. The efficiency-related uncertainties are estimated in data using tag-and-probe techniques. The MC samples used are corrected for differences in these efficiencies between data
and simulation and the associated uncertainties are propagated through the analysis. The lepton energy scale uncertainties are measured in data.  For $\tau_{\text{had}}$ candidates, where the uncertainty is dominated by calorimeter response, this is done by fitting the visible $Z\rightarrow\tau\tau$ mass \cite{ATLAS_Tauperf}.  The systematic uncertainties due to energy resolution have a negligible impact on the result.  Systematic effects from electron- and muon-related uncertainties are smaller in general than those from jets and $\tau_{\text{had}}$ candidates. The soft-scale $\MET$ resolution accounts for low-$\pT$ energy deposits that do not contribute to the clustered energy of physics objects ($e$, $\mu$, $\tau$, jet). The $b$-jet tagging efficiency is measured in data with $\ttbar$ events and has an uncertainty of a few percent, which in turn has a small impact on the prediction of the $\ttbar$ background in the signal region.

The systematic uncertainty on the background from jets misidentified as leptons is estimated  for each type of lepton separately. It is assumed to be uncorrelated with all other uncertainties. 
The uncertainty on the contribution to the background from jets misidentified as $\tau_{\text{had}}$ is dominated by uncertainty in the fraction of quark- and gluon-initiated jets. This accounts for the potential difference between the fraction of quark-initiated jets in the fake-factor measurement region and the analysis signal region, where the fake factor is applied.  Because quark- and gluon-initiated jets can fake $\tau_{\text{had}}$ candidates at different rates, a difference in their ratio between the fake-factor measurement and signal region would bias the fake factors themselves.  The systematic uncertainty is evaluated by varying the ratio of quark- to gluon-initiated jets from half to two times the nominal value, as determined in MC simulation.  The systematic uncertainty for the electron fake factor is determined in a way similar to the $\tau_{\text{had}}$ fake factor, although the compositions of misidentified candidates from jets and photons are varied as opposed to the relative fractions of quark- and gluon-initiated jets. 

The uncertainty on the luminosity ($\pm$2.8\%) derived from beam-separation scans performed in 2012 using the method described in Ref. \cite{Aad:2013ucp} affects the number of signal and simulated background events.

Theoretical uncertainties are estimated for the signal and for all background contributions derived using MC simulation. Uncertainties relating to higher order QCD corrections and MC modeling choices are estimated by varying the renormalization and factorization scales, PDF parameterization and underlying-event model as described in Ref. \cite{ATLAS_HSG4_Paper}. 
The signal samples, generated in QCD LO with \textsc{Pythia8}, are normalized using cross sections computed in NNLO in QCD and NLO in electroweak, but kinematic distributions, such as the Higgs boson $\pt$, are not re-weighted. The HAWK MC program~\cite{hawk2014}, which calculates NLO QCD and NLO electroweak corrections for all the $VH$ processes, is used to evaluate the resulting systematic uncertainties due to kinematic differences.  The impact of the QCD scale choice on the signal acceptance is evaluated in MC simulation before the ATLAS detector simulation is performed, separately for the four analysis channels, by varying the QCD scales in \textsc{Powheg + Pythia8}.

\begin{table}[ht]
	\begin{center}
\caption{
Impact of systematic uncertainties on the expected yields of the signal and/or relevant background(s) as well as the impact on the signal strength $\mu$.  The experimental uncertainties affect the signal prediction and all backgrounds that are determined with MC simulation.  The background model uncertainties affect the prediction of the backgrounds from fake-factor methods. The theoretical uncertainties affect the signal prediction.  Where ranges are given they indicate the variation of the impact on different channels or differences between one-track and multi-track  $\tau_{\text{had}}$ candidates. All values are given before the global fit.}
	\small
	\begin{tabular}{lcc}
	\hline
	    {Source} & {Impact on event yeild [\%]} & Impact on $\mu$ \\ 
	\hline
	    \textbf{{\small Experimental}}  &           & \\
        \hline
        Luminosity                      & $\pm$ 2.8      & $\pm$ 0.30 \\
        Tau identification              & $\pm$ 2--6     & $\pm$ 0.41 \\
        Lepton identification and trigger       & $\pm$ 1--1.8   & $\pm$ 0.15 \\
        $b$-tagging                     & $\pm$2         & $\pm$ 0.16 \\
        $\tau$ energy scale             & $\pm$ 0-- 2.9  & $\pm$ 0.57 \\
        Jet energy scale and resolution & $\pm$ 4        & - \\
	    \MET\ soft scale \& resolution  & $\pm$ 0.1--0.5 & - \\
	\hline
    \textbf{{\small Background Model}}  &               & \\
        \hline
        Modeling of BG from misidentified jets   & $\pm$ 15 -- 38 & $\pm$ 0.72 \\
	\hline
	    \textbf{{\small Theoretical}}   &                & \\ 
	    \hline
        Higher-order QCD corrections    & $\pm$ 2--8\    & $\pm$ 0.26 \\
        Underlying event/parton shower modeling                           & $\pm$ 1--4     & $\pm$ 0.07 \\
        Generator modeling             & $\pm$ 1.4      & $\pm$ 0.05 \\
        EW corrections                  & $\pm$ 2        & $\pm$ 0.06 \\
        PDF                             & $\pm$ 3-4      & $\pm$ 0.18 \\
        ${\cal B}$~($H\to\tau\tau$)             & $\pm$ 3-7      & $\pm$ 0.17 \\
	\hline
	\end{tabular}
	\label{systematics-prefit}
	\end{center}
	\end{table}

\FloatBarrier

\section{Results}
\label{sec:result}

The observed signal strength $\mu$, is determined from a binned global maximum-likelihood fit to the reconstructed Higgs boson candidate mass distributions, with nuisance parameters $\vec{\theta}$, corresponding to the systematic uncertainties. The $M_{\mathrm{2T}}$ distribution is used for the $WH$ topologies and the $M_{\mathrm{MMC}}$ distribution for  the $ZH$ categories. 
For each signal and background process, each nuisance parameter is separately tested to determine whether it affects the $M_{\mathrm{2T}}$ or $M_{\mathrm{MMC}}$ distributions.
For background processes only, the effect of a nuisance parameter on the shape of the distributions is neglected if the difference between the up and down variations of the yield in all bins of the distribution is less than 10\% of the total background  statistical error. Overall systematic uncertainties that differ from the nominal by less than 0.5\% are not considered. The only exception is the treatment of systematic uncertainties due to theoretical aspects, which are fully considered even though they have a small overall impact on the fit.

The expected
numbers of signal and background events in each bin are functions of $\vec{\theta}$.
The test statistic $q_\mu$ is then constructed according to
the profile likelihood ratio: $q_\mu = - 2 \ln [\mathcal{L} (\mu,
\hat{\hat{\vec{\theta}}})/\mathcal{L} (\hat{\mu}, \hat{\vec{\theta}})]$, where the numerator
 $\mathcal{L} (\mu,\hat{\hat{\vec{\theta}}})$ is the conditional maximum likelihood with $\hat{\hat{\vec{\theta}}}$ the value of the nuisance parameters that maximize 
 $\mathcal{L}$ for a given $\mu$ and the denominator $\mathcal{L} (\hat{\mu}, \hat{\vec{\theta}})$ is the unconditional maximum likelihood.
This test statistic is used to measure
the compatibility of the background-only hypothesis with the observed
data and for setting limits derived with the $CL_\mathrm{s}$ method~\cite{Read:2002hq,Cowan:2010js}. To quantify this compatibility, a significance is calculated, giving the probability of obtaining $q_\mu$ if $\mu$ = 1 is the true signal strength.
 
The measured signal strength, normalized to the SM expectation, is $\mu =$ 2.3 $\pm$ 1.6 for $m_H=125\GeV$. The 95\% confidence-level (CL) upper limits for each of  the four channels and their associated signal strengths are  shown in Figure~\ref{Figure:Combfitmodeltests}. The expected and observed significances for each of the four channels are shown in Table~\ref{Table:p0}.

\begin{figure}[ht]
        \centering
   \subfloat{%
	\includegraphics[width=0.49\textwidth]{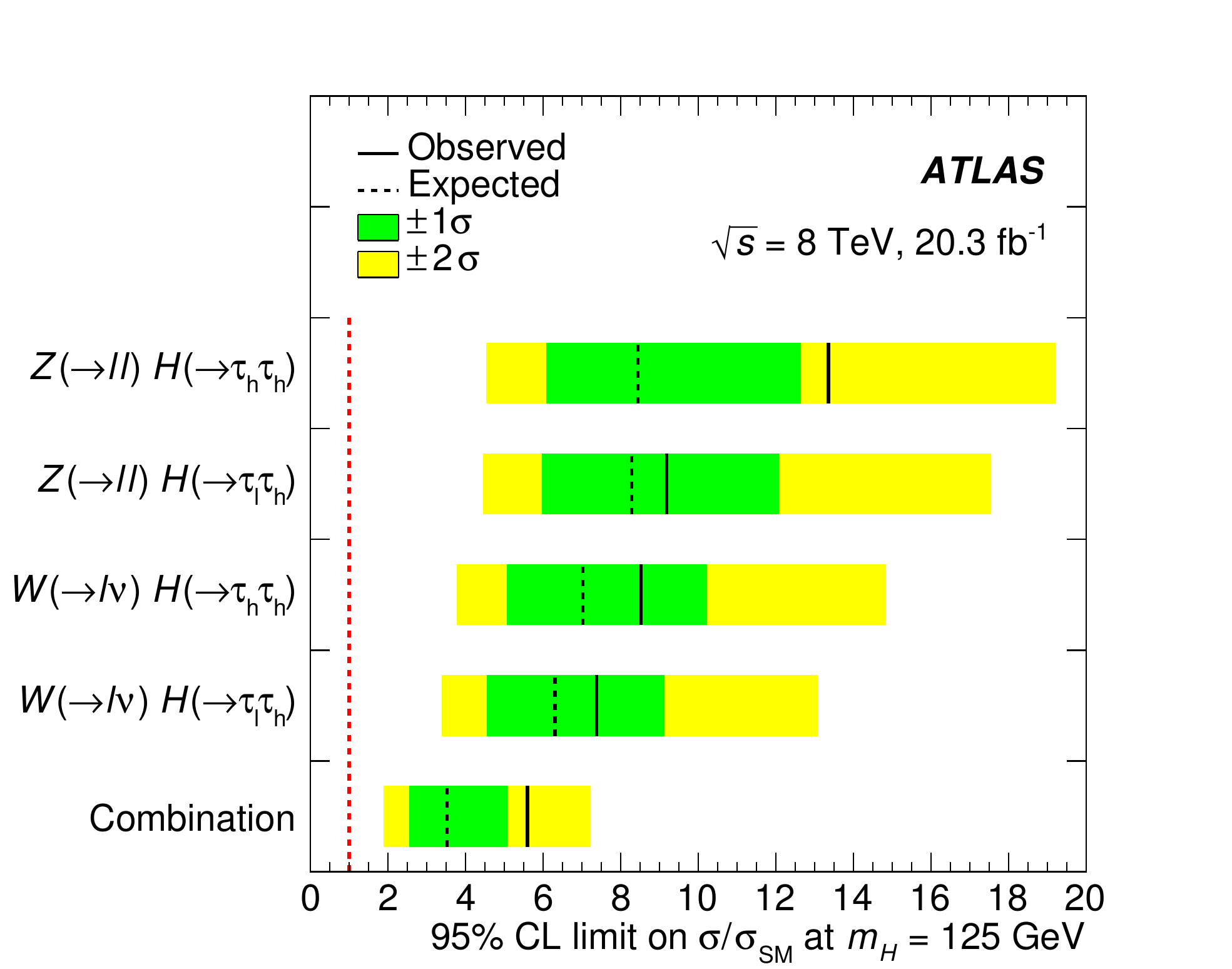}
	}
   \subfloat{%
	\includegraphics[width=0.49\textwidth]{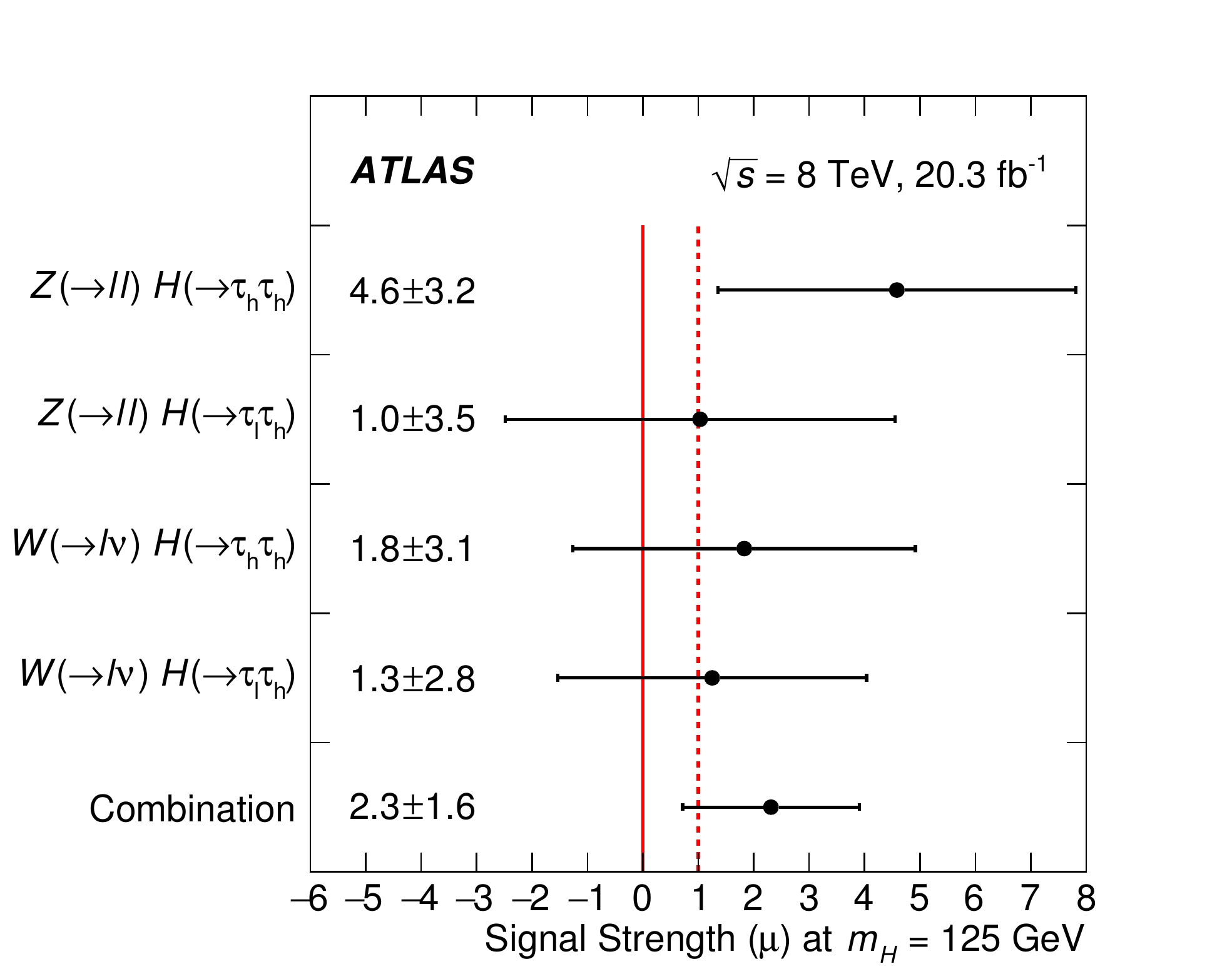}
	}
	\caption{The combined result for the $VH$ channels.  The 95\% CL cross
section limit is shown for each individual channel on the left. The right figure
shows the signal strength in each individual channel, along with the combination.
}
	\label{Figure:Combfitmodeltests}
\end{figure}

\begin{table}[ht]
	\begin{center}
    \caption{The expected and observed significances for the four channels.}
	\small
	\begin{tabular}{lcc}
	    \hline
	    Channel                                                                     & Expected significance & Observed significance       \\
	    \hline
	    $W\rightarrow\mu\nu/e\nu, ~~H\rightarrow\tau_{\text{lep}}\tau_{\text{had}}$ & 0.36  $\sigma$ & 0.44 $\sigma$  \\
        $W\rightarrow\mu\nu/e\nu, ~~H\rightarrow\tau_{\text{had}}\tau_{\text{had}}$ &0.32 $\sigma$  & 0.60 $\sigma$ \\
        $Z\rightarrow\mu\mu/ee, ~~H\rightarrow\tau_{\text{lep}}\tau_{\text{had}}$   & 0.28 $\sigma$   &  0.29 $\sigma$   \\
        $Z\rightarrow\mu\mu/ee, ~~H\rightarrow\tau_{\text{had}}\tau_{\text{had}}$   & 0.32 $\sigma$    &  1.38 $\sigma$   \\
	    \hline
	\end{tabular}
	\label{Table:p0}
	\end{center}
\end{table}

The overall 95\% CL limit on the observed ratio of the cross section to the SM prediction is 5.6 at $m_H=125\GeV$, which is above the expected values of 3.5 if no signal is assumed and 3.7 if signal is included, but is consistent within the uncertainties of the expected limit.   The weaker limit in the data comes mostly from the slight excesses seen  in the two channels with $H\rightarrow \tau_{\text{had}}\tau_{\text{had}}$.

\FloatBarrier

\section{Conclusion}
\label{sec:conclusion}

The analysis presented in this paper, a search for the associated production of the SM Higgs boson with a vector boson where the Higgs boson decays to a pair of tau leptons, is based on 20.3~\ifb\ of LHC proton--proton collisions recorded by the ATLAS experiment at the center-of-mass energy $\sqrt{s}=8 \TeV$. The overall 95\% CL upper limit on the ratio of the observed cross section to the SM predicted cross section, at 5.6, is higher than the expected values of 3.5 if no signal is assumed and 3.7 if signal is included, but is consistent within the statistics and uncertainties of the analysis. 
 The measured signal strength, normalized to the standard model expectation for a Higgs boson of $m_H=125~\GeV$, is $\mu =$ 2.3 $\pm$ 1.6.

\section*{Acknowledgments}


We thank CERN for the very successful operation of the LHC, as well as the
support staff from our institutions without whom ATLAS could not be
operated efficiently.

We acknowledge the support of ANPCyT, Argentina; YerPhI, Armenia; ARC, Australia; BMWFW and FWF, Austria; ANAS, Azerbaijan; SSTC, Belarus; CNPq and FAPESP, Brazil; NSERC, NRC and CFI, Canada; CERN; CONICYT, Chile; CAS, MOST and NSFC, China; COLCIENCIAS, Colombia; MSMT CR, MPO CR and VSC CR, Czech Republic; DNRF, DNSRC and Lundbeck Foundation, Denmark; IN2P3-CNRS, CEA-DSM/IRFU, France; GNSF, Georgia; BMBF, HGF, and MPG, Germany; GSRT, Greece; RGC, Hong Kong SAR, China; ISF, I-CORE and Benoziyo Center, Israel; INFN, Italy; MEXT and JSPS, Japan; CNRST, Morocco; FOM and NWO, Netherlands; RCN, Norway; MNiSW and NCN, Poland; FCT, Portugal; MNE/IFA, Romania; MES of Russia and NRC KI, Russian Federation; JINR; MESTD, Serbia; MSSR, Slovakia; ARRS and MIZ\v{S}, Slovenia; DST/NRF, South Africa; MINECO, Spain; SRC and Wallenberg Foundation, Sweden; SERI, SNSF and Cantons of Bern and Geneva, Switzerland; MOST, Taiwan; TAEK, Turkey; STFC, United Kingdom; DOE and NSF, United States of America. In addition, individual groups and members have received support from BCKDF, the Canada Council, CANARIE, CRC, Compute Canada, FQRNT, and the Ontario Innovation Trust, Canada; EPLANET, ERC, FP7, Horizon 2020 and Marie Sk\l{}odowska-Curie Actions, European Union; Investissements d'Avenir Labex and Idex, ANR, Region Auvergne and Fondation Partager le Savoir, France; DFG and AvH Foundation, Germany; Herakleitos, Thales and Aristeia programmes co-financed by EU-ESF and the Greek NSRF; BSF, GIF and Minerva, Israel; BRF, Norway; the Royal Society and Leverhulme Trust, United Kingdom.

The crucial computing support from all WLCG partners is acknowledged
gratefully, in particular from CERN and the ATLAS Tier-1 facilities at
TRIUMF (Canada), NDGF (Denmark, Norway, Sweden), CC-IN2P3 (France),
KIT/GridKA (Germany), INFN-CNAF (Italy), NL-T1 (Netherlands), PIC (Spain),
ASGC (Taiwan), RAL (UK) and BNL (USA) and in the Tier-2 facilities
worldwide.

\printbibliography

\newpage
\begin{flushleft}
{\Large The ATLAS Collaboration}

\bigskip

G.~Aad$^{\rm 85}$,
B.~Abbott$^{\rm 113}$,
J.~Abdallah$^{\rm 151}$,
O.~Abdinov$^{\rm 11}$,
R.~Aben$^{\rm 107}$,
M.~Abolins$^{\rm 90}$,
O.S.~AbouZeid$^{\rm 158}$,
H.~Abramowicz$^{\rm 153}$,
H.~Abreu$^{\rm 152}$,
R.~Abreu$^{\rm 116}$
% ...
\bigskip
\\
% ...
$^{11}$ Institute of Physics, Azerbaijan Academy of Sciences, Baku, Azerbaijan\\
% ...
$^{85}$ CPPM, Aix-Marseille Universit{\'e} and CNRS/IN2P3, Marseille, France\\
% ...
$^{90}$ Department of Physics and Astronomy, Michigan State University, East Lansing MI, United States of America\\
% ...
$^{107}$ Nikhef National Institute for Subatomic Physics and University of Amsterdam, Amsterdam, Netherlands\\
% ...
$^{113}$ Homer L. Dodge Department of Physics and Astronomy, University of Oklahoma, Norman OK, United States of America\\
% ...
$^{116}$ Center for High Energy Physics, University of Oregon, Eugene OR, United States of America\\
% ...
$^{151}$ Institute of Physics, Academia Sinica, Taipei, Taiwan\\
$^{152}$ Department of Physics, Technion: Israel Institute of Technology, Haifa, Israel\\
$^{153}$ Raymond and Beverly Sackler School of Physics and Astronomy, Tel Aviv University, Tel Aviv, Israel\\
% ...
$^{158}$ Department of Physics, University of Toronto, Toronto ON, Canada\\
% ...
\end{flushleft}


\end{document}